\documentclass{article}
\usepackage{spconf,amsmath,graphicx}
\usepackage{color}
\usepackage{url}
\usepackage{graphicx}
\title{MVNet: Memory Assistance and Vocal Reinforcement Network for Speech Enhancement}
\name{Jianrong Wang$^1$, Xiaomin Li$^1$, Xuewei Li$^1$, Mei Yu$^1$, Qiang Fang$^2$, Li Liu$^{3,*}$\thanks{* Corresponding author}}
\address{$^1$College of Intelligence and Computing, Tianjin University, Tianjin, China\\
$^2$Institute of Linguistics, Chinese Academy of Social Sciences, Beijing, China\\
$^3$Shenzhen Research Institute of Big Data, the Chinese University of Hong Kong, Shenzhen, China}
\begin{document}
%\ninept
%
\maketitle
\vspace{-0.1cm}
\begin{abstract}
Speech enhancement improves speech quality and promotes the performance of various downstream tasks. However, most current speech enhancement work was mainly devoted to improving the performance of downstream automatic speech recognition (ASR), only a relatively small amount of work focused on the automatic speaker verification (ASV) task. In this work, we propose a MVNet consisted of a memory assistance module which improves the performance of downstream ASR and a vocal reinforcement module which boosts the performance of ASV. In addition, we design a new loss function to improve speaker vocal similarity. Experimental results on the Libri2mix dataset show that our method outperforms baseline methods in several metrics, including speech quality, intelligibility, and speaker vocal similarity \textit{et al}.
\end{abstract}
\vspace{-0.1cm}
\section{Introduction}
\vspace{-0.1cm}
The interference of additive noise with speech can seriously reduce the perceptual quality and intelligibility of speech, which increases the difficulty and complexity of speech-related recognition tasks. In some scenarios, the security of algorithms for tasks such as speech recognition and speaker verification can be seriously threatened by noise interference \cite{chen2021real}. Speech enhancement (SE) is an important speech processing task dedicated to improving the perceptual quality as well as the intelligibility of the disturbed speech and to restore the performance of downstream tasks.

A good SE algorithm should obtain the output speech that is closer to the clean speech. And the output speech often has better speech quality and intelligibility than the input speech. In recent years, deep learning methods \cite{9431717} \cite{9411829} \cite{9437977} \cite{hao2021fullsubnet} \cite{karthik2021efficient} were widely applied to SE tasks and achieved good results. Deep learning based methods can be classified into time domain and frequency domain depending on how the input speech is processed. The common practice of time domain methods \cite{pandey2020densely} \cite{chen2020dual} \cite{luo2020dual} \cite{wang2021tstnn} is to map the time domain waveform of noisy speech directly to the time domain waveform of clean speech, through the learned mapping relationship. The frequency domain approach \cite{hao2021fullsubnet} \cite{hu2020dccrn} obtains a mask by inputting the noisy speech spectral features into the network. Then the clean speech is obtained by multiplying the mask and the noisy speech.

\begin{figure*}
\centerline{\includegraphics[width=14cm]{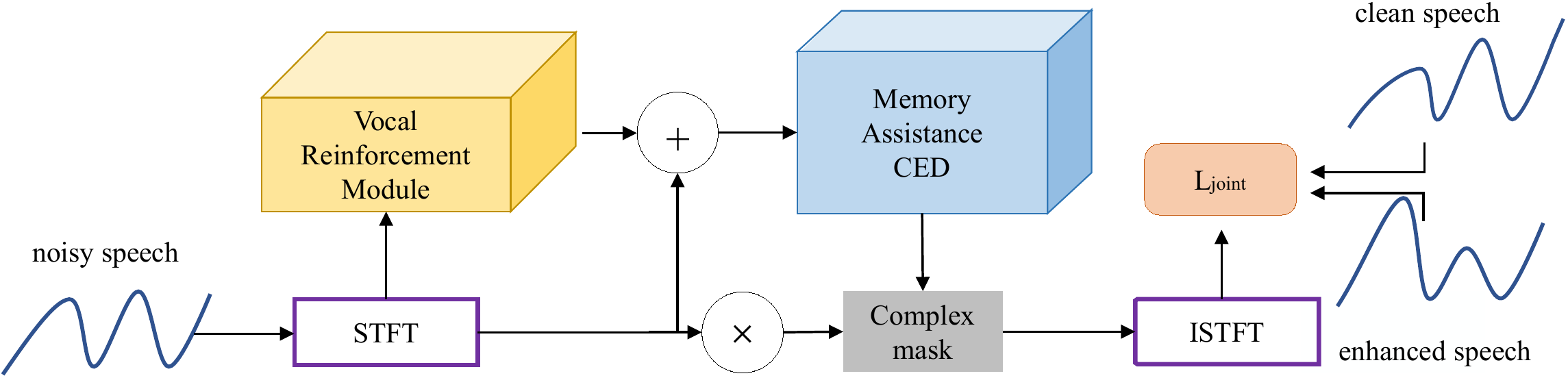}}
\caption{The overall structure of MVNet.}
\label{fig:model_archi}
\end{figure*}

Most of the previous work focused on improving speech quality as a training goal, and the current mainstream metrics are also based on speech quality. Several studies proposed to train SE models directly with speech quality metrics (PESQ and STOI), including quality-net  \cite{fu2019learning}, MetricGAN-u \cite{fu2022metricgan} and hifi-gan \cite{kong2020hifi}. These methods achieved a significant improvement in speech quality. 
However, ASR and ASV pay different attention to speech features. ASR pays more attention to the intelligibility of speech, while ASV pays more attention to speaker vocal similarity. The optimization focus of the two is not consistent. Speech with higher speech quality can have more outstanding performance in the downstream ASR task, while less outstanding in ASV.
Current methods greatly improve speech quality (PESQ and STOI), ignoring the importance of vocal information. However, inconsistent vocals will lead to inconsistencies between speakers and increased distortion, which in turn affects the performance of downstream ASVs. We call this the vocal distortion problem.

PFPL \cite{hsieh2020improving} started to demonstrate the importance of phonetic information. Their work demonstrates that adding the necessary speech information can guarantee speech details as well as speech quality. This provides us with ideas to alleviate the vocal distortion problem. 

In this work, to adapt to both ASR and ASV at the same time, achieve the improvement of speech quality and vocal consistency, we propose a MVNet consisted of a memory assistance module and a vocal feature reinforcement module. Vocal reinforcement module is to extract the vocal information. We consider it important for vocal distortion problem. Memory assistance module is to improve the enhanced performance of the complex network. It reduces the loss from forgetting valid information in long sequences by the network while enhancing the gain from focusing on important information. Besides, we design a similarity joint loss that aims to alleviate vocal distortion problem. The experiments verify that our method can alleviate the vocal distortion problem while further improving the speech quality.

\vspace{-0.1cm}
\section{Related Work}
\vspace{-0.1cm}

% \begin{figure*}[htbp]
% 	\centering
% 	\includegraphics[width=0.83\textwidth,height=2.5in]{./figure/framework.pdf}
% 	\caption{Overview of our FR-PSS with the prior speech.}
% 	\label{fig:pipeline}
% \vspace{-0.3cm}
% \end{figure*}

\subsection{Complex Structure of CRN}
The traditional CRN \cite{tan2018convolutional} network is symmetric. It uses an encoder-decoder architecture in the time-domain, usually with an LSTM layer in the middle to model the temporal dependencies. The encoder-decoder block consists of convolution and deconvolution layers, batch normalization and activation functions.

To improve the performance of convolution in the complex domain, Tan \textit{et al.} \cite{tan2019complex} proposed a one encoder two decoders convolution method. Unlike previous CRN that only targets amplitude mapping in the real domain, this network structure is also capable of modeling phase mapping in the complex domain. Compared with the traditional enhancement model, this structure can enhance the amplitude and phase of the speech at the same time, and the enhanced speech no longer needs to reuse the phase of the noisy speech. However, this one encoder two decoders structure actually divided the input into two channels, the real part and the imaginary part, and processed them as real numbers, which did not strictly follow the operation rules of complex numbers.

The above approaches did not directly utilize the prior knowledge of the magnitude and phase correlations of complex arithmetic. Hu \textit{et al.} provided a complex domain convolution model DCCRN \cite{hu2020dccrn}, which used a complex encoder-decoder combined with a complex LSTM to enhance speech. This network provided the ability to simulate complex multiplication, further enhancing the network's ability to capture the correlation between magnitude and phase. DCCRN has been shown to be effective, and we take it as our baseline model.

\subsection{Speech Feature Information}
% The waveform of the original speech naturally contains phase information. In addition to the above methods, the practice of directly adopting end-to-end processing of noisy speech waveforms to preserve the phase information of the original speech also attracted much attention. The end-to-end method can also reduce the information loss caused by the manual features extraction.

With the research in the signal processing, researchers developed different speech features according to the characteristics of different tasks. Speech feature extraction methods such as MFCC \cite{tiwari2010mfcc} and i-vector \cite{garcia2011analysis} showed value in various speech signal processing tasks such as speech recognition, speaker recognition, and phoneme detection. These feature representations focus on different speech information. A suitable feature representation can strongly promote the performance of a specific task.

Hsieh \textit{et al.} \cite{hsieh2020improving} proposed a perceptual loss (PFPL) for SE task. They pointed out that phonetic feature information is the key to optimizing human perceptual. PFPL first proposed the idea of adding phonetic feature information to the original speech. This self-supervised SE method is based on DCCRN and wav2vec \cite{baevski2020wav2vec}. Their experimental results showed effectiveness of phonetic information. And we take PFPL as another baseline model.

\section{Method}
%\label{sec:pagestyle}
\vspace{-0.1cm}
In this work, we propose a MVNet as shown in Fig. \ref{fig:model_archi}. In general, we extract the speaker vocal features through vocal reinforcement module, and fuse it with the noisy speech spectrum. A complex mask is then estimated by the memory assistance speech enhancement module and multiplied by the noisy spectrum to obtain the enhanced speech. Besides, we use the proposed similarity joint loss to alleviate vocal distortion problem.

% \begin{figure}
% \centerline{\includegraphics[width=8cm]{figures/mvnet.pdf}}
% \caption{The overall structure of MVNet.}
% \label{fig:model_archi}
% \end{figure}

Our method is based on DCCRN which excels in speech quality. We propose the memory assistance module to further improve speech quality and make the model pay more attention to the vocal features. To improve the vocal similarity of speech, we propose the vocal reinforcement module and the similarity joint loss.

\subsection{Memory Assistance}
\begin{figure}
\centerline{\includegraphics[width=0.85\columnwidth]{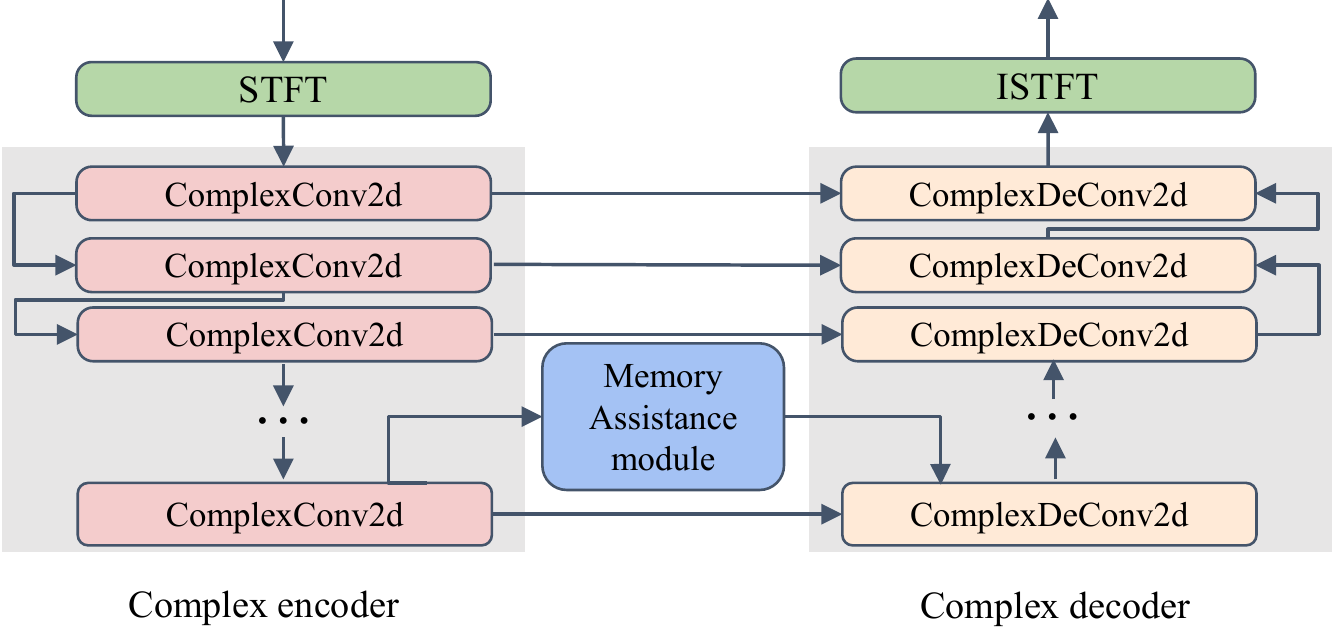}}
\caption{The memory assistance CED structure.}
\label{fig:CED}
\end{figure}

% \subsection{Memory Assistance}
% \begin{figure*}
% \centerline{\includegraphics{figures/CED3.pdf}}
% \caption{The memory assistance CED structure.}
% \label{fig:CED}
% \end{figure*}

In order to make the model further improve the speech quality, and at the same time make it have the ability to pay attention to the vocal features. We propose the memory assistance module under the DCCRN framework, as shown in Fig. \ref{fig:CED}.

% The proposed memory assistance deep complex convolutional encoder-decoder (CED) structure is shown in Fig. \ref{fig:CED}. This part consists of a CED and a memory assistance speech enhancement module.

We use 6 complex convolution blocks and symmetric 6 deconvolution blocks to implement the construction of the encoder-decoder with the number of channels set to \{32, 64, 128, 256, 256, 256\}, where each complex convolution block contains complex Conv2d, complex batch normalization and real-valued PReLU. Each complex Conv2d contains a real conv2d and an imaginary conv2d as in DCCRN \cite{hu2020dccrn}.

% CED follows complex arithmetic rules. The complex convolution filter defines the form of complex values in analog mathematics, as shown in Eq. \ref{equ:complexweight}.

% \begin{equation}
% \text{W}=\text{W}_{r}+j*\text{W}_{i},
% \label{equ:complexweight}
% \end{equation}
% where $\text{W}_{r}$ and $\text{W}_{i}$ are real-valued matrices representing the real and imaginary parts of the complex convolution kernel, which are combined to form the complex convolution kernel $\text{W}$. And $j$ represents the imaginary number. The resulting output is expressed as Eq. \ref{equ:Fout}, where $\text{X}$ is the complex input calculated as $\text{W}$.
% \begin{equation}
% \text{F}_{\text{out}}=(\text{W}_{r}*\text{X}_{r}-\text{W}_{i}*\text{X}_{i}) + j(\text{W}_{i}*\text{X}_{r}-\text{W}_{r}*\text{X}_{i}).
% \label{equ:Fout}
% \end{equation}

\subsubsection{Memory Assistance Module}
The overall framework of the DCCRN model is based on CED, and the speech enhancement is mainly realized by the LSTM with causal modeling ability in between. The LSTM network controls the memory state of information in the long-term transmission process through gates, retains important information and forgets the information that the network considers unimportant. It plays the role of information filtering. Vocal information is a very detailed speech feature that can only be noticed from a global perspective. The core logic of the attention mechanism is the global attention, which can capture the vocal features. But only from a global perspective will weaken some local characteristics of speech. Thus, we combine it with the LSTM to form the memory assistance module, which focuses on the both global features and local details of speech.

\begin{figure}
\centerline{\includegraphics[width=\columnwidth]{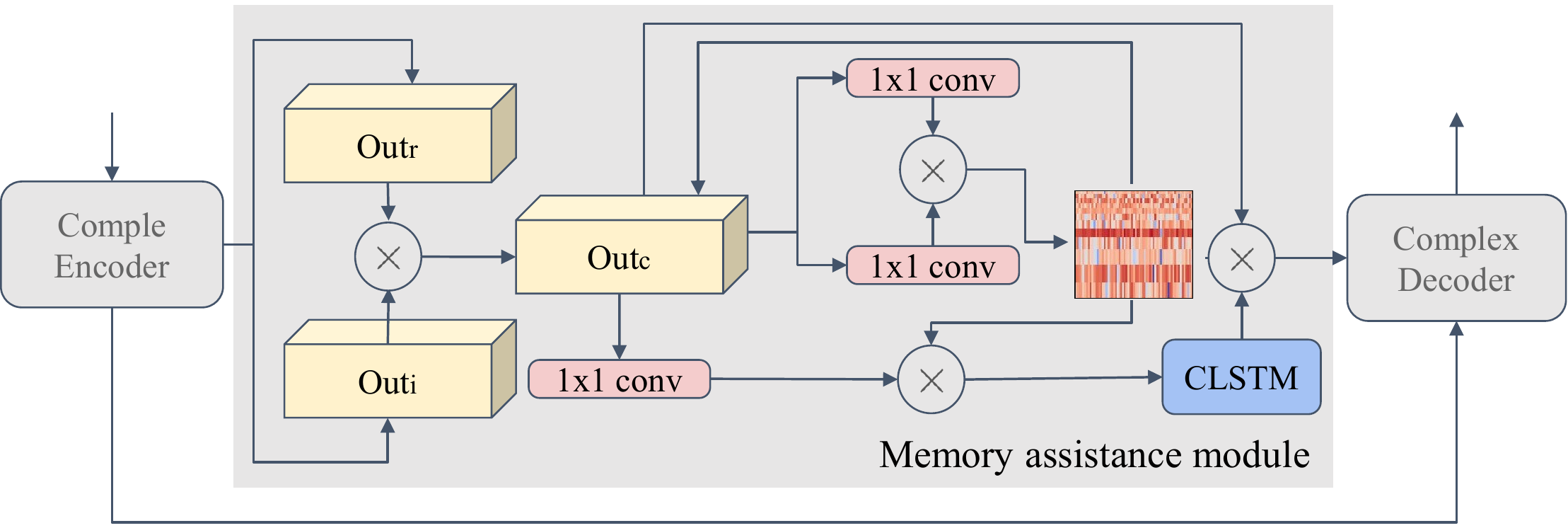}}
\caption{Memory assistance speech enhancement module. $\text{Out}_\text{r}$ and $\text{Out}_\text{i}$ is the real and the imaginary part of the encoder output, respectively. $\text{Out}_\text{c}$ is the fused complex feature map. CLSTM is the complex LSTM layer.} 
\label{fig:MA_module}
\end{figure}

% The core logic of the attention mechanism is to shift from local focus to global focus. Since the calculation of each step in the attention mechanism does not depend on the result of the previous step, it can capture important information from a global perspective. We believe that this character of attention mechanism can reduce the loss of LSTM forgetting effective information in long sequences, while improving the benefits of focusing on important information. We propose a memory assistance module based on the attention mechanism. It assists LSTM to memorize key information, amplify memory gains and reduce forgetting losses.

Placing the attention before LSTM can amplify the memory ability, improves the memory ability of LSTM for global vocal characteristics. If it is placed in the back, LSTM will forgot this information. At this time, the global characteristics of this information will be destroyed, and the global attention will not be able to pay attention to this information. Placing the attention in the back aggravates the forgetting ability. Thus, the final vocal reinforcement is as Fig. \ref{fig:MA_module}.

We utilize the features on the crisscross path to achieve global attention through two loops while controlling the memory consumption. Module collects contextual information in both horizontal and vertical directions to enhance the expressiveness of feature maps. As shown in Fig. \ref{fig:MA_module}, the noisy speech is passed through the complex encoder to obtain the feature maps of both the real and imaginary parts. The feature maps is fused and send into three Conv1ds. The horizontal and vertical attention map is obtained from the first two Conv1ds and then passed back to the input to obtain the global attention map. The global attention map and the output of the third Conv1d are concatenated and fed into the complex LSTM. The output is concatenated with the output of the encoder and fed to the decoder for further processing.

\subsection{Vocal Reinforcement}
Vocal feature is an important factor affecting the distortion degree of the final enhanced speech. When the vocal features of the enhanced speech and the clean speech are quite different, the speech sounds lack of uniform speaker characteristics, and it does not sound like the original speaker. That causes vocal distortion problem.

The problem of missing vocal characteristics in the process of enhancing speech can be considered from two perspectives. One is that the model does not have the ability to discover such characteristics, and the other is that the optimization direction of the model does not care about this. To improve the vocal similarity of speech from these two perspectives, we propose the vocal reinforcement module and similarity joint loss.

\subsubsection{Vocal Reinforcement Module}
For the first perspective, our solution is to explicitly add vocal features to the network, which is the direct idea of our vocal reinforcement module. 

\begin{figure}
\centerline{\includegraphics[width=\columnwidth]{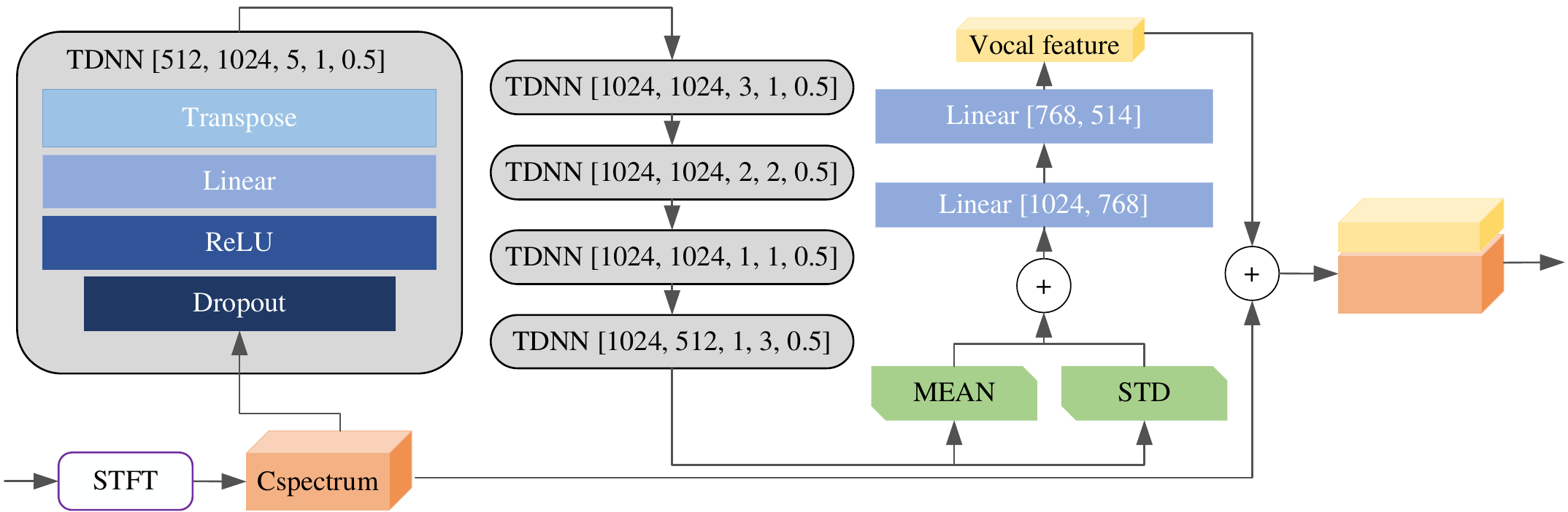}}
\caption{Vocal reinforcement module.}
\label{fig:vocal_extracter}
\end{figure}

The way of combining MFCC with TDNN \cite{ji2011prediction} is a common way to obtain speaker representation in ASV, which has a strong expressive ability for vocal features. But MFCC is a compact speech representation. Since it uses mel filter to ignore the dynamics and distribution of speech energy, it still loses some speech details in essence \cite{9195238}. It is a coarse-grained speech representation. Therefore, we adopt STFT (short-time Fourier transform) to obtain the spectral representation of speech and preserve the temporal information of speech. Combined with TDNN, a fine-grained vocal feature extraction method suitable for speech enhancement is realized. ASV directly uses the speech representation obtained by MFCC and TDNN to do the recognition task. But speech enhancement requires more than just vocal information. We therefore combine the obtained representation as an auxiliary feature with the spectrum obtained by STFT as the input to CED. This forms our vocal reinforcement module.

% We believe that the vocal feature is an important factor affecting the distortion degree of the final enhanced speech. When the vocal features of the enhanced speech and the clean speech are quite different, the speech sounds lack of uniform speaker characteristics, and it does not sound like the original speaker. That causes vocal distortion problem.

The proposed vocal reinforcement module is shown in Fig. \ref{fig:vocal_extracter}. The spectrum obtained by STFT is sent to 5 TDNNs connected in sequence, the first four output 1024 channels, the last collapses the channels to 512, and the average and standard deviation of the last TDNN`s output are calculated and connected to the original output, and then go through two linear layers in turn to get the vector about the vocal features. We fuse the feature vector and the original input into Memory Assistance CED.

\subsubsection{Similarity Joint Loss}
As mentioned above, to improve the vocal consistency, there are two perspectives. Designing a new loss function is from the second, changing the direction of model optimization.

The complete information of the speech signal is jointly represented by the amplitude and the phase, and the phase contains more detailed information of the speech. In the previous speech enhancement models, SI-SNR \cite{luo2019conv} was mostly used as the loss function. Although SI-SNR takes into account the vector direction of speech, the calculation process still depends on the signal amplitude. The cosine similarity has a stronger constraint on the consistency of the vector direction. In order to make the model pay more attention to the vector direction, we introduce the cosine similarity to our loss function. The ability of the model to improve the vocal consistency is enhanced by strengthening the constraint of the loss function on the consistency of the vector direction.

The proposed similarity joint loss is to make some improvements on the basis of the loss function of SI-SNR. We take the additive inverse of SI-SNR in our $\mathcal{L}_{\text{SI-SNR}}$, so that larger calculated results indicate less ideal separation. $\mathcal{L}_{\text{SI-SNR}}$ is defined as Eq. \ref{equ:SI-SNR loss}:
% \begin{equation}
%     \text{SI\_SNR} =10 \log_{10}\left(\frac{\left\|\frac{(<\hat{s}, s>\cdot s)}{\|s\|_{2}^{2}}\right\|_{2}^{2}}{\left\|\hat{s}-s_{target}\right\|_{2}^{2}}\right)
% \label{equ:SI-SNR}
% \end{equation}

\begin{equation}
\left\{\begin{array}{l}
\mathbf{s}_{\text {target }}=\frac{\langle\hat{\mathbf{s}, \mathbf{s}\rangle \mathbf{s}}}{\|\mathbf{s}\|^{2}} \\
\mathbf{e}_{\text {noise }}=\hat{\mathbf{s}}-\mathbf{s}_{\text {target }} \\
\mathcal{L}_\text {SI-SNR}=10 \log _{10} \frac{\left\|\mathbf{s}_{\text {target }}\right\|^{2}}{\left\|\mathbf{e}_{\text {noise }}\right\|^{2}}
\end{array}\right.
\label{equ:SI-SNR loss}
\end{equation}

where $<\cdot ,\cdot >$ represents the dot product of two vectors,  $\| \cdot \|^{2}$ is the euclidean norm (L2 norm), $\mathbf{s}$ is the clean speech, and $\hat{\mathbf{s}}$ means the enhanced speech. SI-SNR is commonly used in papers. 

In addition to $\mathcal{L}_{\text{SI-SNR}}$, we propose to use cosine similarity to improve the speaker vocal consistency. The similarity loss is defined as Eq. \ref{equ:similarity loss}:
\begin{equation}
\mathcal{L}_{\text{smi}}=\alpha \log_{10} (1-\text{cos}_{\text{smi}}(\hat{s}, s)+\delta),
\label{equ:similarity loss}
\end{equation}
where the hyperparameter $\alpha$ is the scaling factor, which we set to 100.

The value range of the cosine similarity function $\text{cos}_{\text{smi}}(\cdot , \cdot)$ is [-1, 1]. %and the higher the value, the more similar the two speeches are. 
We take the -$\text{cos}_{\text{smi}}$, so that the higher the calculation result, the more dissimilar the two speeches are. We add a constant number 1 to fix the range in [0,2]. $\delta$ is an extremely small number used to avoid zero values. 
%This satisfies the domain requirement of the logarithmic function. 
We smooth the change of the curve through a logarithmic function, so that $\mathcal{L}_{\text{smi}}$ has a consistent change trend with $\mathcal{L}_{\text{SI-SNR}}$. Finally, we combine these two functions to propose our similarity joint loss $\mathcal{L}_{\text{joint}}$, defined as Eq. \ref{equ:similarity joint loss}:
\begin{equation}
\mathcal{L}_{\text{joint}}=\mathcal{L}_{\text{SI-SNR}}+\mathcal{L}_{\text{smi}}.
\label{equ:similarity joint loss}
\end{equation}

%In the experiment, we use the enhanced speech and clean speech to calculate the loss after ISTFT, and then update the network weights with $\mathcal{L}_{\text{joint}}$ to improve the enhancement performance.

\subsection{Training Target}
Our training target is to obtain a complex ratio mask(CRM) \cite{williamson2015complex} to estimate clean speech. We adopt the method of signal estimation, that is, the noisy signal and the estimated mask are directly multiplied to obtain the enhanced signal. We improve the performance of the model by minimizing the $\mathcal{L}_{\text{joint}}$ between the enhanced and the clean speech.

\vspace{-0.1cm}
\section{Experiment and Result}
\vspace{-0.1cm}
%\label{sec:typestyle}
\subsection{Setup}
\subsubsection{Dataset}
In our experiments, we use the Librispeech \cite{panayotov2015librispeech} as the clean data, which has 1252 speakers, each speaking for about 25 minutes, for a total of 478 hours of speech duration. The noise data in the experiment comes from the noise dataset WHAM! \cite{wichern2019wham}, which consists of real ambient noises.

% , and the sound samples are collected from coffee shops, restaurants, and bars in the San Francisco Bay Area and publicly available

We mix the Lirispeech and WHAM! datasets in the same way as the LibriMix \cite{cosentino2020librimix}, resulting in a training set with 921 speakers for a total of 364 hours, and a validation set and a test set with 40 speakers for a total of 5.4 hours. The dataset SNR we get from the mix is between -15bB and 5dB.

\subsubsection{Evaluation Metrics}
The evaluation of our experiments is based on several general metrics of speech quality, including Perceptual Evaluation of Speech Quality (PESQ) \cite{rix2001perceptual}, Short-Time Objective Intelligibility (STOI) \cite{taal2011algorithm}, the predicted Mean Opinion Score of signal distortion (CSIG) \cite{hu2007evaluation}, background noise distortion (CBAK) \cite{hu2007evaluation}, overall quality (COVL) \cite{hu2007evaluation}, the scale-invariant signal-to-noise ratio improvement (SI-SNRi) \cite{vincent2006performance}, segmental SNR (segSNR) \cite{quackenbush1986objective} and SIMI (a measure of vocal similarity). SIMI is the proposed new metric to measure the degree of vocal distortion, which is calculated by the speaker recognition algorithm provided by Deep-speaker \cite{li2017deep}. The higher the score, the higher the probability that the speaker will be judged to be the same in the ASV task.

\subsubsection{Setup and Baseline}
We sample waveforms at 16kHz, and set the window length and number of hops to 25 ms and 6.25 ms, respectively. The FFT length is 512. We use Adam optimizer. The initial learning rate is set to 0.001, and when the validation loss increases, the learning rate decreases by 0.5. We train for 200 epochs and record the top PESQ ranked model parameters as our best model for related experiments. For fairness, we run the official codes of baseline models (PFPL and DCCRN) with the same training configuration as ours for comparison.

\subsection{Ablation Study for Memory Assistance}
We propose the memory assistance module to further improve speech quality and make the model pay more attention to the vocal features. 

% \begin{table}
% \caption{Ablation study for memory assistance. $\text{Ours}_{\text{ma}}$ and $\text{Ours}_{\text{bma}}$ represents the result of placing the memory assistance module before the LSTM layer and after the LSTM, respectively.
% DCCRN represents the results obtained without memory assistance.}
% \centering
% \setlength{\tabcolsep}{2.5mm}
% \begin{tabular}{|l|l|l|l|l|}
% \hline
%     Metric& noisy& $\text{Ours}_{\text{bma}}$& $\text{DCCRN}$& $\text{Ours}_{\text{ma}}$\\
%     \hline
%     PESQ&1.18&2.52&2.65&\textbf{2.70}\\
%     STOI&0.54&0.81&0.84&\textbf{0.87}\\
%     SIMI&0.36&0.39&0.43&\textbf{0.46}\\ \hline
% \end{tabular}
% \label{table:Memory assistance module}
% \end{table}

\vspace{-0.6cm}
\begin{table}[htb]
  \caption{Ablation study for memory assistance. $\text{Ours}_{\text{ma}}$ and $\text{Ours}_{\text{bma}}$ represents the result of placing the memory assistance module before the LSTM layer and after the LSTM, respectively.
DCCRN represents the results obtained without memory assistance.}
  \setlength{\tabcolsep}{2.5mm}
  \label{table:Memory assistance module}
  \centering
  \resizebox{.48\textwidth}{!}{
 \begin{tabular}{ccccc}
		\hline
		Metric& noisy& $\text{Ours}_{\text{bma}}$& $\text{DCCRN}$& $\text{Ours}_{\text{ma}}$\\
		\hline
		PESQ&1.18&2.52&2.65&\textbf{2.70}\\
        STOI&0.54&0.81&0.84&\textbf{0.87}\\
        SIMI&0.36&0.39&0.43&\textbf{0.46}\\
		\hline
	\end{tabular}}
\end{table}

From Table \ref{table:Memory assistance module}, it can be seen that memory assistance module has the best results when placed before LSTM. Memory assistance module can amplify the importance of effective information before the LSTM forgets some information, so that LSTM continues to amplify those effective information. When the memory assistance module is placed behind LSTM, the degree of forgetting of the LSTM will be aggravated, thereby reducing the performance of the model.

Memory assistance module outperforms DCCRN in PESQ, STOI and SIMI. This indicates its ability to further improve speech quality while empowering the model to focus on vocal features.

\subsection{Ablation Study for Vocal Reinforcement}
To improve the vocal similarity of speech, we propose the vocal reinforcement module and the similarity joint loss. We compare with PFPL(with phonetic information) and DCCRN(without any speech information). Results are shown in Table \ref{table:Vocal featurese module}, our results are significantly better than the above methods. 
\begin{table}
\caption{Ablation study for vocal reinforcement. $\text{Ours}_{\text{vr}}$ is the model with only the vocal reinforcement. $\text{Ours}_{\text{MVL}}$ is the MVNet with both memory assistance and vocal reinforcement.}
\centering
\setlength{\tabcolsep}{1.5mm}
\begin{tabular}{ccccccc}
    \hline
    Metric& noisy& $\text{DCCRN}$& PFPL&$\text{Ours}_{\text{vr}}$ &$\text{Ours}_{\text{MVL}}$\\ 
    \hline
    PESQ&1.18&2.65&2.71&\textbf{2.90}&2.88\\
    STOI&0.54&0.84&0.78&\textbf{0.91}&\textbf{0.91}\\
    SIMI&0.36&0.43&0.51&\textbf{0.52}&\textbf{0.52}\\
    SegSNR&2.07&6.49&5.58&5.88&\textbf{6.55}\\
    CSIG&2.02&2.10&2.37&\textbf{2.47}&2.44\\
    $\text{SI-SNR}_\text{i}$&-&6.98&6.27&9.88&\textbf{9.97}\\
    \hline
\end{tabular}
\label{table:Vocal featurese module}
\end{table}

PFPL explicitly added the additional information (phonetic information) to the DCCRN. This additional information contained more speech details, which made the PESQ, SIMI and CSIG of PFPL higher than those of DCCRN. However, PFPL did not modify the original CED structure of DCCRN, and its model did not have the ability to adapt to this fine-grained information. Thus, its STOI, SegSNR and $\text{SI-SNR}_\text{i}$ were degraded. Compared with DCCRN and PFPL, both $\text{Ours}_{\text{vr}}$ and $\text{Ours}_{\text{MVL}}$ have better SIMI and CSIG scores, which prove that the vocal reinforcement can improve the vocal consistency. Since $\text{Ours}_{\text{vr}}$ lacks the memory assistance module in $\text{Ours}_{\text{MVL}}$, the local details (SegSNR) and overall performance ($\text{SI-SNR}_\text{i}$) of $\text{Ours}_{\text{vr}}$ are worse than $\text{Ours}_{\text{MVL}}$.
%$\text{Ours}_{\text{vr}}$ alleviates the vocal distortion problem, improving CSIG and $\text{SI-SNR}_\text{i}$ while achieving higher PESQ and STOI than baseline models. $\text{Ours}_{\text{MVL}}$ achieves higher SegSNR and $\text{SI-SNR}_\text{i}$ with higher vocal consistency(SIMI) while maintaining higher speech quality(PESQ and STOI). 
\subsection{Comprehensive Evaluation}

\begin{table*}[!h]
\caption{Comprehensive evaluation.}
\centering
\setlength{\tabcolsep}{1.55mm}
\begin{tabular}{ccccccccc}
\hline
    & PESQ& STOI& CSIG& CBAK& CVOL& SIMI& SegSNR& $\text{SI-SNR}_\text{i}$\\ \hline
    noisy&1.18&0.54&2.02&1.99&1.75&0.36&2.07&-\\
    $DCCRN$&2.65&0.84&2.10&2.11&1.91&0.43&6.49&9.04\\
    PFPL&2.71&0.78&2.37&2.45&2.12&0.51&5.58&8.34\\ \hline
    $\text{Ours}_{\text{ma}}$&2.70&0.87&2.26&2.34&2.08&0.46&6.37&10.61\\
    $\text{Ours}_{\text{vr}}$&\textbf{2.90}&\textbf{0.91}&\textbf{2.47}&\textbf{2.60}&\textbf{2.61}&\textbf{0.52}&5.88&11.94\\
    $\text{Ours}_{\text{MVL}}$&2.88&\textbf{0.91}&2.44&2.59&2.27&\textbf{0.52}&\textbf{6.55}&\textbf{12.04} \\\hline
\end{tabular}
\label{table:Comprehensive evaluation}
\end{table*}
We comprehensively evaluate the performance of our method on various metrics, as shown in Table \ref{table:Comprehensive evaluation}. Our model outperforms the baseline models in all metrics and lower distortion can be guaranteed while maintaining higher speech quality.

\vspace{-0.5cm}
\section{Conclusion}
\vspace{-0.3cm}
In this work, we propose the MVNet consisted of a memory assistance module and a vocal reinforcement module. Memory assistance module is proposed to further improve the speech quality while making the model focus more on vocal features.  Vocal reinforcement module explicitly introduces vocal features to improve the speaker vocal similarity. Besides, we design a similarity joint loss, which aims to improve the speaker vocal consistency. Experiments verify that the MVNet can further improve speech quality while maintaining the increase in speaker similarity and the decrease in speech distortion, which correspond to the concerns of the ASR and ASV tasks, respectively. In the future, we will continue to explore what affects speech enhancement performance.

%\vfill\pagebreak

\bibliographystyle{IEEEbib}
\bibliography{refs}

\end{document}